\documentclass[11pt]{article}
\usepackage{multicol}
\usepackage{graphicx}
\usepackage{amssymb}
\usepackage{amsthm}
\title{\bf Superposed Coherent and Squeezed Light}
\author{Fesseha Kassahun\footnote{Email address: fesseha@phys.aau.edu.et} \\
                \footnotesize{Department of Physics, Addis Ababa University, P. O. Box 33761, Addis Ababa, Ethiopia}}
                \date{\footnotesize{(Submitted on 14 January 2012)}}

\makeatother

\begin{document}
\maketitle
\begin{abstract}
We first calculate the mean photon number and quadrature variance of superposed coherent and squeezed light, following a procedure of analysis based on combining the Hamiltonians and using the usual definition for the quadrature variance of superposed light beams. This procedure of analysis leads to physically unjustifiable mean photon number of the coherent light and quadrature variance of the superposed light. We then determine both of these properties employing a procedure of analysis based on superposing the Q functions and applying a slightly modified definition for the quadrature variance of a pair of superposed light beams. We find the expected mean photon number of the coherent light and the quadrature variance of the superposed light. Moreover, the quadrature squeezing of the superposed output light turns out to be equal to that of the superposed cavity light.
\end{abstract}
\hspace*{9.5mm}Keywords: Q function, Mean photon number, Quadrature Squeezing, Superposition
\vspace*{5mm}
\begin{multicols}{2}
\section*{1. Introduction}
 \vspace*{-2mm}
It has been established that a one-mode subharmonic generator produces light with a maximum quadrature squeezing of $50\%$ below the coherent-state level [1-4]. In addition, some authors [5-7] have studied the statistical and squeezing properties of superposed coherent and squeezed light, produced by harmonic and subharmonic generations in the same cavity. These authors have carried out their analysis by combining the pertinent Hamiltonians and applying the usual definition for quadrature variance. But such procedure leads to physically unjustifiable mean photon number of the coherent light and quadrature variance of the superposed coherent and squeezed light.

In order to see clearly the problems connected with the mean photon number of the coherent light and quadrature variance of the superposed coherent and squeezed light, we undertake a procedure of analysis based on combining the Hamiltonians and using the usual definition for the quadrature variance of the superposed light.
From the results of our analysis, we discover that the presence of the squeezed light has some effect on the mean photon number of the coherent light. This must certainly be due to the procedure of analysis we have employed. Moreover, we find that the quadrature variance of the superposed coherent and squeezed light is equal to that of the squeezed light only. In other words, the coherent light has no contribution to the quadrature variance of the superposed light. The origin of this problem must be the use of the usual definition for the quadrature variance of the superposed light.

In order to avoid the aforementioned problems, we calculate the mean photon number and quadrature variance of the superposed coherent and squeezed light, applying a procedure of analysis based on superposing the Q functions and slightly modifying the usual definition for the quadrature variance of the superposed light. We also determine the quadrature squeezing of the superposed cavity and output light.

\section*{2. The Hamiltonian combining \hspace*{8mm} procedure}
\subsection*{2.1 The mean photon number}

We seek to obtain the mean photon number and quadrature variance of superposed coherent and squeezed light, produced by harmonic and subharmonic generations, in a cavity coupled to a vacuum reservoir via a single-port mirror. We consider the case in which the coherent and squeezed light beams have the same frequency. The process of harmonic generation is described by the Hamiltonian
\begin{equation}\label{1}
\hat{H}_{c}=i\varepsilon_{1}(\hat{a}^{\dagger}-\hat{a}),
\end{equation}
where $\varepsilon_{1}$ is proportional to the amplitude of the coherent light driving the cavity mode. And the process of subharmonic generation is described by the Hamiltonian
 \begin{equation}\label{2}
\hat{H}_{s}={i\varepsilon_{2}\over 2}(\hat{a}^{2}-\hat{a}^{\dagger 2}),
\end{equation}
with $\varepsilon_{2}$ being proportional to the amplitude of the coherent light pumping the nonlinear crystal.
The analysis of the superposed coherent and squeezed light is usually carried out employing the Hamiltonian [5-7]
\begin{equation}\label{3}
\hat{H}=i\varepsilon_{1}(\hat{a}^{\dagger}-\hat{a})+{i\varepsilon_{2}\over 2}(\hat{a}^{2}-\hat{a}^{\dagger 2}),
\end{equation}
which is the sum of the Hamiltonians given by Eqs. (\ref{1}) and (\ref{2}).

We will calculate the expectation values of only normally-ordered operators pertaining to a cavity mode coupled to a vacuum reservoir. We then note that the noise operator associated with the vacuum reservoir has no effect on the dynamics of the cavity mode operators. Hence upon dropping this noise operator, the quantum Langevin equation for the operator $\hat{a}$ can be written as
\begin{equation}\label{4}
{d\hat{a}(t)\over dt}=-{1\over 2}\kappa\hat{a}(t)-i[\hat{a}(t),\hat{H}],
\end{equation}
where $\kappa$ is the cavity damping constant. Therefore, on account of (\ref{4}) and (\ref{3}), we have
\begin{equation}\label{5}
{d\hat{a}(t)\over dt}=-{1\over 2}\kappa\hat{a}(t)-\varepsilon_{2}\hat{a}^{\dagger}(t)+\varepsilon_{1}.
\end{equation}
Now applying the relation
\begin{equation}\label{6}
{d\over dt}\langle\hat{a}(t)\hat{a}(t)\rangle=\bigg\langle{d\hat{a}(t)\over dt}\hat{a}(t)\bigg\rangle+\bigg\langle\hat{a}(t){d\hat{a}(t)\over dt}\bigg\rangle
\end{equation}
along with Eq. (\ref{5}), we get
\begin{eqnarray}\label{7}
{d\over dt}\langle\hat{a}(t)\hat{a}(t)\rangle\hspace*{-3mm}&=\hspace*{-3mm}&-\kappa\langle\hat{a}^{2}(t)\rangle-2\varepsilon_{2}
\langle\hat{a}^{\dagger}(t)\hat{a}(t)\rangle\nonumber\\&&
+2\varepsilon_{1}\langle\hat{a}(t)\rangle-\varepsilon_{2}.
\end{eqnarray}
Moreover, using the relation
\begin{equation}\label{8}
{d\over dt}\langle\hat{a}^{\dagger}(t)\hat{a}(t)\rangle=\bigg\langle{d\hat{a}^{\dagger}(t)\over dt}\hat{a}(t)\bigg\rangle+\bigg\langle\hat{a}^{\dagger}(t){d\hat{a}(t)\over dt}\bigg\rangle
\end{equation}
together with Eq. (\ref{5}), we find
\begin{eqnarray}\label{9}
{d\over dt}\langle\hat{a}^{\dagger}(t)\hat{a}(t)\rangle\hspace*{-3mm}&=\hspace*{-3mm}&-\kappa\langle\hat{a}^{\dagger}(t)\hat{a}(t)\rangle
-\varepsilon_{2}\langle\hat{a}^{2}(t)\rangle\nonumber\\&&
-\varepsilon_{2}\langle\hat{a}^{\dagger 2}(t)\rangle+\varepsilon_{1}
\langle\hat{a}(t)\rangle\nonumber\\&&
+\varepsilon_{1}\langle\hat{a}^{\dagger}(t)\rangle.
\end{eqnarray}

The steady-state solutions of Eqs. (\ref{7}) and (\ref{9}) have the form
\begin{equation}\label{10}
\langle\hat{a}^{2}\rangle=-b\langle\hat{a}^{\dagger}\hat{a}\rangle+a\langle\hat{a}\rangle-{b\over 2},
\end{equation}
\begin{equation}\label{11}
 \langle\hat{a}^{\dagger}\hat{a}\rangle=-{b\over 2}\langle\hat{a}^{2}\rangle-{b\over 2} \langle\hat{a}^{\dagger 2}\rangle+{a\over 2}\langle\hat{a}\rangle+{a\over 2}\langle\hat{a}^{\dagger}\rangle,
\end{equation}
where
\begin{equation}\label{12}
a=2\varepsilon_{1}/\kappa
\end{equation}
and
\begin{equation}\label{13}
b=2\varepsilon_{2}/\kappa.
\end{equation}
In addition, the steady-state solution of the expectation value of Eq. (\ref{5}) turns out to be
\begin{equation}\label{14}
\langle\hat{a}\rangle=-b\langle\hat{a}^{\dagger}\rangle+a.
\end{equation}
We also note that
\begin{equation}\label{15}
\langle\hat{a}^{\dagger}\rangle=-b\langle\hat{a}\rangle+a.
\end{equation}
Upon adding and subtracting Eq. (\ref{15}) to and from Eq. (\ref{14}), we arrive at
\begin{equation}\label{16}
\langle\hat{a}\rangle{\pm}\langle\hat{a}^{\dagger}\rangle={a{\pm}a\over 1+b}.
\end{equation}
It then follows that
\begin{equation}\label{17}
\langle\hat{a}\rangle=\langle\hat{a}^{\dagger}\rangle={a\over{b+1}}.
\end{equation}

Furthermore, with the aid of (\ref{17}), Eqs. (\ref{10}) and (\ref{11}) can be rewritten as
\begin{equation}\label{18}
\langle\hat{a}^{2}\rangle=-b\langle\hat{a}^{\dagger}\hat{a}\rangle+{2a^{2}-b(1+b)\over 2(1+b)},
\end{equation}
\begin{equation}\label{19}
 \langle\hat{a}^{\dagger}\hat{a}\rangle=-b\langle\hat{a}^{2}\rangle+{a^{2}\over{1+b}}.
\end{equation}
Following the same procedure as the one leading to the result given by (\ref{17}), we obtain from Eqs. (\ref{18}) and (\ref{19}) that
\begin{equation}\label{20}
\langle\hat{a}^{2}\rangle={a^{2}\over{(1+b)^{2}}}+{b\over{2(b^{2}-1)}}
\end{equation}
and
\begin{equation}\label{21}
\langle\hat{a}^{\dagger}\hat{a}\rangle={a^{2}\over{(1+b)^{2}}}+{b^{2}\over{2(1-b^{2})}}.
\end{equation}
Now on account of (\ref{12}) and (\ref{13}), we see that
\begin{equation}\label{22}
\langle\hat{a}^{\dagger}\hat{a}\rangle={4\varepsilon_{1}^{2}\over(\kappa+2\varepsilon_{2})^{2}}+{2\varepsilon_{2}^{2}
\over{\kappa^{2}-4\varepsilon_{2}^{2}}}.
\end{equation}
Evidently the second term in Eq. (\ref{22}) is the mean photon number of the squeezed light. However, the first term does not represent the mean photon number of the coherent light. The mean photon number of the coherent light is just the first term with $\varepsilon_{2}=0$. We clearly see that the presence of the squeezed light has some effect on the mean photon number of the coherent light. There cannot be any physically valid justification for this effect. The origin of the problem connected with the first term in Eq. (\ref{22}) must certainly be the procedure of analysis we used, which is based on combining the Hamiltonians.

\subsection*{2.2 Quadrature variance}
We next wish to calculate the quadrature variance for the superposed coherent and squeezed light. The variance of the quadrature operators
\begin{equation}\label{23}
\hat{a}_{+}=\hat{a}^{\dagger}+\hat{a}
\end{equation}
and
\begin{equation}\label{24}
\hat{a}_{-}=i(\hat{a}^{\dagger}-\hat{a}),
\end{equation}
 for the superposed light, is usually defined by
\begin{equation}\label{25}
(\Delta a_{\pm})^{2}=1+\langle:\hat{a}_{\pm},\hat{a}_{\pm}:\rangle.
\end{equation}
This can be put in the form
\begin{eqnarray}
(\Delta a_{\pm})^{2}\hspace*{-3mm}&=\hspace*{-3mm}&1+2\langle\hat{a}^{\dagger}\hat{a}\rangle
{\pm}\langle\hat{a}^{2}\rangle{\pm}\langle\hat{a}^{\dagger 2}\rangle\nonumber\\&&{\mp}\langle\hat{a}\rangle^{2}{\mp}\langle\hat{a}^{\dagger}\rangle^{2}
-2\langle\hat{a}^{\dagger}\rangle\langle\hat{a}\rangle.
\end{eqnarray}
Now employing Eqs. (\ref{17}), (\ref{20}), and (\ref{21}), one can readily verify that
\begin{equation}\label{26}
(\Delta a_{\pm})^{2}=1{{\mp}{b\over{1{\pm}b}}}
\end{equation}
and in view of (\ref{13}), we have
\begin{equation}\label{27}
(\Delta a_{\pm})^{2}=1{{\mp}{2\varepsilon_{2}\over{\kappa{\pm}2\varepsilon_{2}}}}.
\end{equation}
This is just the quadrature variance of the squeezed light alone [4]. Contrary to our expectation, the coherent light has no contribution to the quadrature variance of the superposed light. We maintain standpoint that the problems connected with the first term in Eq. (\ref{22}) and the quadrature variance described by Eq. (\ref{27}) could be resolved by carrying out the pertinent analysis based on superposing the Q functions, instead of combining the Hamiltonians, and by slightly modifying the usual definition for the quadrature variance of superposed light beams.

\section*{3. The Q function superposing \hspace*{9mm}procedure}
\subsection*{3.1 The Q function}
In this section we first calculate the Q functions for the coherent and squeezed light beams. Then using these results, we determine the Q function for the superposed coherent and squeezed light.

\subsubsection*{3.1.1 The coherent light Q function}
We now seek to obtain the Q function for the coherent light produced by harmonic generation. To this end, upon setting $\varepsilon_{2}$=0 in Eq. (\ref{5}), we have
\begin{equation}\label{29}
{d\hat{a}(t)\over dt}=-{1\over 2}\kappa\hat{a}(t)+\varepsilon_{1}.
\end{equation}
The solution of this equation can be written as
\begin{equation}\label{30}
\hat{a}(t)=a(1-e^{-\kappa t/2}) +\hat{a}'(t),
 \end{equation}
where
\begin{equation}\label{31}
\hat{a}'(t)=\hat{a}(0)e^{-\kappa t/2}
\end{equation}
and $a$ is given by (\ref{12}).
Then we easily get
 \begin{equation}\label{32}
{d\hat{a}'(t)\over dt}=-{1\over 2}\kappa\hat{a}'(t)
\end{equation}
and with the coherent light considered to be initially in a vacuum state, we see that
\begin{equation}\label{33}
\langle\hat{a}'(t)\rangle=0.
\end{equation}
The Q function for a single-mode light can be expressed as
\begin{equation}\label{34}
Q(\alpha^{*},\alpha,t)={1\over\pi^{2}}\int d^{2}z\phi_{a}(z,t)e^{z^{*}\alpha-z\alpha^{*}},
\end{equation}
in which
\begin{equation}\label{35}
\phi_{a}(z,t)=\langle e^{-z^{*}\hat{a}(t)}e^{z\hat{a}^{\dagger}(t)}\rangle
\end{equation}
is the antinormally-ordered characteristic function.
With the aid of the identity
\begin{equation}\label{36}
e^{\hat{A}}e^{\hat{B}}=e^{\hat{A}+\hat{B}+[\hat{A},\hat{B}]},
\end{equation}
Eq. (\ref{35}) can be put in the form
\begin{equation}\label{37}
\phi_{a}(z,t)=e^{-z^{*}z/2}\langle e^{z\hat{a}^{\dagger}(t)-z^{*}\hat{a}(t)}\rangle.
\end{equation}
Now on substituting (\ref{30}) into Eq. (\ref{37}) and observing that the operator $\hat{a}'(t)$ is a Gaussian variable with a vanishing mean, we have
\begin{eqnarray}\label{38}
\phi_{a}(z,t)\hspace*{-3mm}&=\hspace*{-3mm}&exp[-z^{*}z+a(z-z^{*})(1-e^{-\kappa t/2})
\nonumber\\&&+{1\over 2}z^{2}\langle\hat{a}'^{2}(t)\rangle+{1\over 2}z^{* 2}\langle\hat{a}'^{\dagger 2}(t)\rangle\nonumber\\&&
-z^{*}z\langle\hat{a}'^{\dagger}(t)\hat{a}'(t)\rangle].
\end{eqnarray}
Furthermore, applying Eq, (\ref{31}) and considering the coherent light to be initially in a vacuum state, one easily gets
\begin{equation}\label{39}
\langle\hat{a}'^{2}(t)\rangle=\langle\hat{a}'^{\dagger}(t)\hat{a}'(t)\rangle=0.
\end{equation}
Therefore, in view of this result, Eq. (\ref{38}) takes at steady state the form
\begin{equation}\label{40}
\phi_{a}(z)=exp[-z^{*}z+a(z-z^{*})].
\end{equation}
Finally, upon introducing (\ref{40}) into Eq. (\ref{34}) and carrying out the integration, the coherent light Q function is found to be
\begin{equation}\label{41}
Q(\alpha^{*},\alpha)={1\over\pi}exp[-\alpha^{*}\alpha+a(\alpha+\alpha^{*})-a^{2}].
\end{equation}

\subsection*{3.1.2 The squeezed light Q function}

We next proceed to determine the Q function for the squeezed light produced by subharmonic generation. Thus upon setting $\varepsilon_{1}=0$ in Eqs. (\ref{5}), (\ref{20}), and (\ref{21}), we have
\begin{equation}\label{42}
{d\hat{a}(t)\over dt}=-{1\over 2}\kappa\hat{a}(t)-\varepsilon_{2}\hat{a}^{\dagger}(t),
\end{equation}
 \begin{equation}\label{43}
\langle\hat{a}^{2}\rangle={b\over{2(b^2-1)}},
\end{equation}
and
\begin{equation}\label{44}
\langle\hat{a}^{\dagger}\hat{a}\rangle=-{b^{2}\over{2(1-b^2)}}.
\end{equation}
 Using Eq. (\ref{42}) and assuming the squeezed light to be initially in a vacuum state, one can readily verify that $\langle\hat{a}(t)\rangle=0$. Now on realizing that the operator $\hat{a}(t)$ is a Gaussian variable with a vanishing mean and taking into account (\ref{43}) together with (\ref{44}), Eq. (\ref{37}) can be put at steady state in the form
 \begin{equation}\label{45}
 \phi_{a}(z)=exp[-a_{1}z^{*}z+a_{2}(z^{2}+z^{* 2})/2],
 \end{equation}
in which
\begin{equation}\label{46}
a_{1}=1+{b^{2}\over{2(1-b^{2})}},
\end{equation}
\begin{equation}\label{47}
a_{2}={b\over{2(b^{2}-1)}}.
\end{equation}
Hence introducing (\ref{45}) into Eq. (\ref{34}) and carrying out the integration, the squeezed light Q function is found to be
\begin{eqnarray}\label{48}
Q(\alpha^{*},\alpha)\hspace*{-3mm}&=\hspace*{-3mm}&{[u^{2}-v^{2}]^{1\over 2}\over\pi}exp[-u\alpha^{*}\alpha\nonumber\\&&
+v(\alpha^{2}+\alpha^{* 2})/2],
\end{eqnarray}
where
\begin{equation}\label{49}
u={1-b^{2}/2\over{1-b^{2}/4}}
\end{equation}
and
\begin{equation}\label{50}
v=-{b/2\over{1-b^{2}/4}}.
\end{equation}

\subsection*{3.1.3 The superposed light Q function}
We finally seek to calculate the Q function for the superposed coherent and squeezed light. This Q function is expressible as [8]
\begin{eqnarray}\label{51}
Q(\alpha^{*},\alpha)\hspace*{-3mm}&=\hspace*{-3mm}&{1\over\pi}\int d^{2}\beta d^{2}\gamma Q_{c}(\beta^{*}, \alpha-\gamma)\nonumber\\&&
\times Q_{s}(\gamma^{*},\alpha-\beta)
exp(-\alpha^{*}\alpha-\beta^{*}\beta\nonumber\\&&
-\gamma^{*}\gamma
+\alpha^{*}\beta+\alpha\beta^{*}+\alpha^{*}\gamma+\alpha\gamma^{*}\nonumber\\&&
-\beta^{*}\gamma-\beta\gamma^{*}),
\end{eqnarray}
where $Q_{c}(\beta^{*},\alpha-\gamma)$ and $Q_{s}(\gamma^{*},\alpha-\beta)$ represent the Q functions for the coherent and squeezed light beams. With the aid of (\ref{41}) and (\ref{48}), one can put Eq. (\ref{51}) in the form
\begin{eqnarray}\label{52}
Q(\alpha^{*},\alpha)\hspace*{-3mm}&=\hspace*{-3mm}&{[u^{2}-v^{2}]^{1\over 2}\over\pi^{3}}\int d^{2}\beta d^{2}\gamma
\exp[-\alpha^{*}\alpha\nonumber\\&&
-\beta^{*}\beta-\gamma^{*}\gamma+[\alpha^{*}+(u-1)\gamma^{*}\nonumber\\&&
-v\alpha]\beta+a\beta^{*}+(\alpha^{*}-a)\gamma\nonumber\\&&
+(1-u)\alpha\gamma^{*}+a\alpha-a^{2}\nonumber\\&&
+v(\alpha^{2}+\beta^{2}+\gamma^{* 2})/2],
\end{eqnarray}
so that on carrying out the integration, we readily arrive at
\begin{eqnarray}\label{53}
Q(\alpha^{*},\alpha)\hspace*{-3mm}&=\hspace*{-3mm}&{A\over\pi}\exp[-u\alpha^{*}\alpha
+v(\alpha^{2}+\alpha^{* 2})/2\nonumber\\&&
+a(u-v)(\alpha+\alpha^{*})],
\end{eqnarray}
in which
\begin{equation}\label{54}
A=[u^{2}-v^{2}]^{1\over 2}exp[a^{2}(v-u)].
\end{equation}

\section*{3.2 The mean photon number}
We now proceed to calculate the mean photon number of the superposed coherent and squeezed light. We recall that the expectation value of an operator function $A(\hat{a}^{\dag},\hat{a})$ can be written as
\begin{equation}\label{55}
\langle\hat{A}\rangle=\int d^{2}\alpha Q(\alpha^{*},\alpha)A_{a}(\alpha^{*},\alpha),
\end{equation}
where $A_{a}(\alpha^{*},\alpha)$ is the c-number function
corresponding to $\hat{A}(\hat{a}^{\dag},\hat{a})$ in the antinormal order.
Now applying (\ref{53}), one can put Eq. (\ref{55}) in the form
\begin{eqnarray}\label{56}
\langle\hat{A}\rangle\hspace*{-3mm}&=\hspace*{-3mm}&{A\over\pi}\int\alpha d^{2}\exp[-u\alpha^{*}\alpha+v(\alpha^{2}+\alpha^{* 2})/2\nonumber\\&&
+a(u-v)(\alpha+\alpha^{*})]A_{a}(\alpha^{*},\alpha).
\end{eqnarray}
Now using the fact that $A_{a}=\alpha^{*}\alpha-1$, the mean photon number can be written as
\begin{eqnarray}\label{57}
\langle\hat{a}^{\dag}\hat{a}\rangle\hspace*{-3mm}&=\hspace*{-3mm}&{A\over\pi}\int d^{2}\alpha \exp[-u\alpha^{*}\alpha+v(\alpha^{2}
+\alpha^{* 2})/2\nonumber\\&&
+a(u-v)(\alpha+\alpha^{*})]\alpha^{*}\alpha-1,
\end{eqnarray}
so that on carrying out the integration, there follows
\begin{eqnarray}\label{58}
\langle\hat{a}^{\dag}\hat{a}\rangle=a^{2}+{b^{2}\over 2(1-b^{2})}.
\end{eqnarray}
Finally, in view of (\ref{12}) and (\ref{13}), we see that
\begin{equation}\label{59}
\langle\hat{a}^{\dagger}\hat{a}\rangle={4\varepsilon_{1}^{2}\over\kappa^{2}}+{2\varepsilon_{2}^{2}
\over{\kappa^{2}-4\varepsilon_{2}^{2}}}.
\end{equation}
As expected this is the sum of the mean photon number of the coherent light and that of the squeezed light.

We next wish to calculate the mean photon number of the output light. The output mode operator $\hat{a}_{out}$ can be written in terms of the cavity mode operator $\hat{a}$ and the input mode operator $\hat{a}_{in}$ as
\begin{equation}\label{60}
\hat{a}_{out}=\sqrt{\kappa}\hat{a}-\hat{a}_{in},
\end{equation}
where $\kappa$ is the cavity damping constant. When calculating the expectation values of only normally-ordered output operators, with the cavity mode coupled to a vacuum reservoir, one can use the relation
\begin{equation}\label{61}
\hat{a}_{out}=\sqrt{\kappa}\hat{a}.
\end{equation}
Then the mean photon number of the output light, defined by
\begin{equation}\label{62}
\overline{n}_{out}=\langle\hat{a}^{\dagger}_{out}\hat{a}_{out}\rangle,
\end{equation}
is expressible as
\begin{equation}\label{63}
\overline{n}_{out}=\kappa\langle\hat{a}^{\dagger}\hat{a}\rangle.
\end{equation}
We then see that the mean photon number of the output light is just $\kappa$ times that of the cavity light.

\section*{3.3 Quadrature squeezing}

We finally discus the quadrature squeezing of the superposed coherent and squeezed light. We define the variance of the quadrature operators
\begin{equation}\label{64}
\hat{a}_{+}=\hat{a}^{\dagger}+\hat{a}
\end{equation}
and
\begin{equation}\label{65}
\hat{a}_{-}=i(\hat{a}^{\dagger}-\hat{a}),
\end{equation}
for a pair of superposed light beams, by
\begin{equation}\label{66}
(\Delta a_{\pm})^{2}=2+\langle:\hat{a}_{\pm},\hat{a}_{\pm}:\rangle.
\end{equation}
This can be rewritten as
\begin{eqnarray}\label{67}
(\Delta a_{\pm})^{2}\hspace*{-3mm}&=\hspace*{-3mm}&2+2\langle\hat{a}^{\dagger}\hat{a}\rangle{\pm}\langle\hat{a}^{2}\rangle{\pm}\langle\hat{a}^{\dagger 2}\rangle\nonumber\\&&
{\mp}\langle\hat{a}\rangle^{2}{\mp}\langle\hat{a}^{\dagger}\rangle^{2}
-2\langle\hat{a}^{\dagger}\rangle\langle\hat{a}\rangle.
\end{eqnarray}

We next proceed to determine the expectation value of the operator $\hat{a}^{2}$. This expectation value can be written employing (\ref{53}) as
\begin{eqnarray}\label{68}
\langle\hat{a}^{2}\rangle\hspace*{-3mm}&=\hspace*{-3mm}&{A\over\pi}\int d^{2}\alpha \exp[-u\alpha^{*}\alpha+v(\alpha^{2}+\alpha^{* 2})/2\nonumber\\&&
+a(u-v)(\alpha+\alpha^{*})\alpha^{2}.
\end{eqnarray}
Hence on carrying out the integration, we readily get
\begin{equation}\label{69}
\langle\hat{a}^{2}\rangle=a^{2}+{b\over{2(b^{2}-1)}}.
\end{equation}
Moreover, following the same procedure, one can easily verify that
\begin{equation}\label{70}
\langle\hat{a}\rangle=a.
\end{equation}
Now on account of Eq. (\ref{67}) along with  (\ref{58}), (\ref{69}), and (\ref{70}), the quadrature variance of the superposed coherent and squeezed light turns out to be
\begin{equation}\label{71}
(\Delta a_{\pm})^{2}=2{\mp}{b\over 1{\pm}b}.
\end{equation}
We observe that the quadrature variance given by (\ref{71}) is the sum of the quadrature variance of the coherent light and that of the squeezed light.

One usually calculates the quadrature squeezing of a pair of superposed light beams relative to the quadrature variance of a single coherent light beam. But now this does not appear to be a correct procedure. We then assert that the quadrature squeezing of a pair of superposed light beams must be calculated relative to the quadrature variance of a pair of superposed coherent light beams. Evidently, employing (\ref{67}) one easily finds the quadrature variance of a pair of superposed coherent light beams to be two. We see from Eq. (\ref{71}) that the squeezing of the superposed coherent and squeezed light occurs in the plus quadrature. We then define the quadrature squeezing of the superposed  light by
\begin{equation}\label{72}
S={2-(\Delta a_{+})^{2}\over 2}
\end{equation}
and in view of (\ref{71}), we see that
\begin{equation}\label{73}
S={b\over 2(1+b)}.
\end{equation}
This is just half of the quadrature squeezing of the squeezed light.

Finally, we seek to determine the quadrature squeezing of the output light. We define the variance of the quadrature operators
\begin{equation}\label{74}
\hat{a}^{out}_{+}=\hat{a}_{out}^{\dagger}+\hat{a}_{out}
\end{equation}
and
\begin{equation}\label{75}
\hat{a}^{out}_{-}=i(\hat{a}_{out}^{\dagger}-\hat{a}_{out}),
\end{equation}
for the superposed coherent and squeezed output light, by
\begin{equation}\label{76}
(\Delta a^{out}_{\pm})^{2}=2\kappa+\langle:\hat{a}^{out}_{\pm},\hat{a}^{out}_{\pm}:\rangle,
\end{equation}
with $2\kappa$ being the quadrature variance of a pair of superposed coherent output light.
On account of (\ref{61}) and (\ref{66}), one can put Eq. (\ref{76}) in the form
\begin{equation}\label{77}
(\Delta a^{out}_{\pm})^{2}=\kappa(\Delta a_{\pm})^{2}.
\end{equation}
We now realize that the quadrature variance of the output light is just $\kappa$ times that of the cavity light.
Moreover, we define the quadrature squeezing of the superposed coherent and squeezed output light by
\begin{equation}\label{78}
S^{out}={2\kappa-(\Delta a^{out}_{+})^{2}\over 2\kappa}.
\end{equation}
Hence on account of Eq. (\ref{77}), we see that
\begin{equation}\label{79}
S^{out}={2-(\Delta a_{+})^{2}\over 2},
\end{equation}
which is exactly the same as the result described by Eq. (\ref{72}).
We then see that the quadrature squeezing of the output light is equal to that of the cavity light.

\section*{4. Conclusion}
 Our calculation of the mean photon number and quadrature variance of the superposed coherent and squeezed light, following a procedure of analysis based on combining the Hamiltonians and using the usual definition for the quadrature variance of superposed light beams, leads to physically unjustifiable mean photon number of the coherent light and quadrature variance of the superposed light. We have then determined both properties employing a procedure of analysis based on superposing the Q functions and applying a slightly modified definition for the quadrature variance of a pair of superposed light beams. We have found the usual mean photon number of the coherent light and the quadrature variance turned out be, as expected, the sum of the quadrature variance of the coherent light and that of the squeezed light. In addition, our analysis shows that the quadrature squeezing of the output light is exactly same as that of the cavity light. It is also perhaps worth mentioning that the presence of the coherent light leads to an increase in the mean photon number and to a decrease in the quadrature squeezing.

\vspace*{5mm}
\noindent
{\bf References}
\vspace*{3mm}

\noindent
[1] G.J. Milburn  and D.F. Walls, Phys. Rev. \hspace*{5.8mm}A 27, 392 (1983).\newline
[2] M.J. Collet and C.W. Gardiner, Phys. Rev. \hspace*{5.8mm}A 30, 1386 (1984).\newline
[3] J. Anwar and M.S. Zubairy, Phys. Rev. \hspace*{5.8mm}A 45, 1804 (1992).\newline
[4] B. Daniel and K. Fesseha, Opt. Commun. \hspace*{5.8mm}151, 384 (1998).\newline
[5] Berihu Teklu, Opt. Commun. 261, 310 \hspace*{5.8mm}(2006).\newline
[6]Tewodros Y. Darge and  Fesseha Kassahun, \hspace*{5.8mm}PMC Physics B, 1 (2010).\newline
[7] Misrak Getahun, PhD Dissertation (Addis \hspace*{5.8mm}Ababa University, 2009).\newline
[8] Fesseha Kassahun, Fundamentals of Quan-\hspace*{5.8mm}tum Optics (Lulu Press Inc., North Car-\hspace*{5.8mm}olina, 2008).\newline

\end{multicols}
\end{document}